\documentclass[twocolumn,showpacs,showkeys,preprintnumbers,amsmath,amssymb,nofootinbib]{revtex4}
\usepackage{graphicx}

\newcommand{\no}{\nonumber}
\newcommand{\stau}{\tilde\tau}
\newcommand{\neu}{\tilde\chi}
\newcommand{\nef}{\tilde\chi_1^0}
\newcommand{\Gstau}{\Gamma_{\stau_1}}
\newcommand{\lam}{\lambda}
\newcommand{\sig}{\sigma}
\newcommand{\cm}{{\cal M}}
\newcommand{\im}{M(\neu\neu)}

\begin{document}

\preprint{KEK-TH-1057}
\title{Correlated decays of pair-produced scalar taus} 

\author{Kaoru Hagiwara$^{1,2,}$}
\email{kaoru.hagiwara@kek.jp}
\author{Kentarou Mawatari$^{1,}$}
\email{kentarou@post.kek.jp}
\affiliation{%
 $^1$Theory Group, KEK, Tsukuba 305-0801, Japan\\
 $^2$Dept.\ of Particle and Nuclear Physics,
     Graduate Univ.\ for Advanced Studies, Tsukuba 305-0801, Japan}  

\author{David Rainwater}
\email{rain@pas.rochester.edu}
\affiliation{%
 Dept.\ of Physics and Astronomy, Univ.\ of Rochester, Rochester,
 NY 14627, USA}

\author{Tim Stelzer}
\email{tstelzer@uiuc.edu}
\affiliation{%
 Dept.\ of Physics, Univ.\ of Illinois at Urbana-Champaign, 1110
 West Green Street, Urbana, Illinois 61801, USA}

\date{\today}

\begin{abstract}
We study the quantum mechanical correlation between two identical
neutralinos in the decays of minimal supersymmetric standard model
 (MSSM) scalar tau (stau) pair produced in 
$e^+e^-$ annihilation.  Generally, the decay products of scalar
(spinless) particles are not correlated.  We show that a correlation
between two neutralinos appears near pair production threshold, due to
a finite stau width and mixing of the staus and/or neutralinos, and
because the neutralinos are Majorana. Because the correlation is
significant only in a specific kinematical configuration, it can be
observed only in supersymmetric models where the neutralino momenta can
be kinematically reconstructed, such as in models with $R$-parity
violation. 
\end{abstract}

\maketitle


\section{Introduction}

Neutralinos are mixtures of the spin-1/2 superpartners of the neutral
gauge and Higgs bosons in supersymmetric extensions to the Standard
Model.  The lightest one ($\nef$) is generally considered to be the
lightest supersymmetric particle (LSP), which is stable in $R$-parity
conserving scenarios.  Signals of supersymmetry (SUSY) are expected to be
 observable at
 hadron~\cite{Dawson:1983fw,Beenakker:1996ed,Beenakker:1999xh,Hinchliffe:1996iu} or future $e^+e^-$
colliders~\cite{Ellis:1982xd,Weiglein:2004hn}.

In the minimal supersymmetric Standard Model (MSSM), neutralinos are
Majorana fermions whose antiparticles are themselves.  If SUSY (and
the associated neutralinos) is observed in nature, the determination
of their properties under $CP$ conjugation, Majorana vs Dirac, will be
a key issue: it is intimately related to the dimension of
SUSY, e.g. $N=1$ or $N\geq 2$ SUSY~\cite{Fox:2002bu}.

Many studies of the Majorana nature of neutralinos have been
performed \cite{Choi:2001ww,CPT,Aguilar-Saavedra:2003hw,Choi:2003fs}.
Reference \cite{CPT} observed that the $CP$ and $CPT$ 
properties of the final state determine the Majorana nature of
neutralinos in the process
$e^+e^-\to\nef\tilde{\chi}_i^0\to\ell^+\ell^-\nef\nef$ ($i=2,3,4$).
Selectron pair production in $e^-e^-$ scattering, $e^-e^-\to\tilde
e^-\tilde e^-$, would also be of great utility, as it could proceed
only on account of $t$-channel Majorana neutralino
exchange~\cite{Aguilar-Saavedra:2003hw}.  In addition, Majorana
effects in two- or three-body decays of neutralinos are known to
exist~\cite{Choi:2003fs}.

In this article, we study the Majorana nature of neutralinos in
lighter scalar tau (stau) pair production and subsequent decay into
$\tau$ lepton (tau) plus LSP neutralino in $e^+e^-$ annihilation,
\begin{equation}
 e^+e^-\to\stau_1^+\stau_1^-\to\tau^+\tau^-\nef\nef \; ,
\end{equation}
via the quantum mechanical correlation between two identical
neutralinos which exists only if they are Majorana fermions.  That a
correlation should exist should be evident simply due to proper
treatment of the Fermi statistics of the final-state neutralinos, but
it is always ignored: conventional wisdom says that the correlation
between decay products of scalar (spinless) particles, such as staus,
is absent.  Because of this, little attention in the literature has
been given to the expected Majorana effect, even though the process we
consider here is among the most promising to study SUSY particles and
many studies of it have been
performed~\cite{Nojiri:1994it,Nojiri:1996fp} for exploration of other
aspects.  Naturally, no correlation is possible once a spinless
particle is separated from the other particles at a macroscopic
space-time distance.  In other words, any interference effect
disappears in the limit of long lifetime of the spinless particles.
We should hence consider the finite stau width, and investigate the
region near pair production threshold.

In addition to the finite width, the left-right mixing of the staus
and/or the gaugino-higgsino mixing of the neutralinos are necessary
for the correlation to be significant.  This is because in the
massless neutralino limit, same-helicity neutralinos are produced only
when the taus have the same helicity, which occurs only when the
$\stau_L$-$\stau_R$ mixing and/or the neutralino mixing are
significant.  We will discuss this in detail in the following section,
and simply note in passing that significant interference effects are
expected in $\tilde e_L^{\pm}\tilde e_R^{\mp}$ and
$\tilde\mu_L^{\pm}\tilde\mu_R^{\mp}$ pair production processes, which
quantitative study will be reported elsewhere.

The article is organized as follows. 
Section \ref{secmix} gives a brief introduction of the mass eigenstates
of the staus ($\stau_{1,2}$) and the gaugino-higgsino mixing of the
neutralinos.  The $\stau_1$-$\tau$-$\nef$ coupling is also given.
In Sec.~\ref{sechel}, we present scattering amplitudes for the process
$e^+e^-\to\stau^+_1\stau^-_1\to\tau^+\tau^-\nef\nef$ and define the
kinematical variables relevant to our analysis.
Section \ref{sectot} gives the total cross sections around the $\stau_1$
pair production threshold.
In Sec.~\ref{secdif}, we study in detail the kinematical correlations
due to identical-particle interference effects and compare them to the
Dirac (noninterference) case.
Section \ref{secsum} is devoted to a summary and discussions. 


\section{Stau and neutralino mixing}
\label{secmix}

In this section, we briefly introduce the staus' left-right mixing and
the neutralinos' gaugino-higgsino mixing.  We also give the
$\stau_1$-$\tau$-$\nef$ coupling, where $\stau_1$ is defined to be the
lighter stau mass eigenstate after $L$-$R$ mixing.

Because of the non-negligible tau Yukawa coupling, mixing occurs
between the weak eigenstates $\stau_L$ and $\stau_R$ to form mass
eigenstates $\stau_1$ and $\stau_2$ ($m_{\stau_1}<m_{\stau_2}$) as
\begin{equation}
  \begin{pmatrix}
   \stau_L \\ \stau_R
  \end{pmatrix}
 =\begin{pmatrix}
    \cos\theta_{\stau} & \sin\theta_{\stau} \\
   -\sin\theta_{\stau} & \cos\theta_{\stau}
  \end{pmatrix} 
  \begin{pmatrix}
   \stau_1 \\ \stau_2
  \end{pmatrix} \, .   
\end{equation}

Neutralinos are mass eigenstates of the neutral gauginos, $\tilde B$
and $\tilde W_3$, and the neutral higgsinos, $\tilde H_{d}^0$ and
$\tilde H_{u}^0\,$.  The mass matrix in the
$X=(\tilde{B},\tilde{W}_3,\tilde{H}^0_d,\tilde{H}^0_u)$ basis is
\begin{equation}
  M=\begin{pmatrix}
     M_1 & 0 & -m_Zs_Wc_{\beta} & m_Zs_Ws_{\beta}  \\
     0 & M_2 & m_Zc_Wc_{\beta}  & -m_Zc_Ws_{\beta} \\
     -m_Zs_Wc_{\beta}  & m_Zc_Wc_{\beta} & 0 & -\mu \\
     m_Zs_Ws_{\beta}   & -m_Zc_Ws_{\beta} & -\mu & 0 
   \end{pmatrix}   
\end{equation}
with the abbreviations $s_W=\sin\theta_W$, $c_W=\cos\theta_W$,
$s_{\beta}=\sin\beta$, and $c_{\beta}=\cos\beta$.  Here, $M_1$ and
$M_2$ are the gaugino masses, $\mu$ is the higgsino mass, and
$\tan\beta=\langle H^0_u\rangle/\langle H^0_d\rangle$ is the ratio of
the vacuum expectation values of the two Higgs doublets.  The mass
eigenstates are given by
\begin{equation}
X_i=U_{ij}\,\neu_j^0 \; , 
\end{equation}
where $U$ diagonalizes the above mass matrix as
\begin{equation}
U^TM\,U={\rm diag}(m_{\nef},m_{\neu^0_2},m_{\neu^0_3},m_{\neu^0_4}) \; .
\end{equation}
Here,
$0\leq m_{\nef}\leq m_{\neu^0_2}\leq m_{\neu^0_3}\leq m_{\neu^0_4}\,$. 

The decay $\stau_1\to\tau\nef$ depends on both the left-right stau
mixing ($\theta_{\stau}$) and the neutralino mixing ($U_{ij}$). 
The $\stau_1$-$\tau$-$\nef$ coupling can be expressed as  
\begin{align}
  {\cal L}
 =\overline\psi_{\nef} \, (a_-P_-+a_+P_+) \, 
  \psi_{\tau}\,\phi_{\stau_1}^*  \, + \, {\rm h.c.} \, ,
\end{align} 
with the chiral-projection operators
$P_{\pm}=\frac{1}{2}(1\pm\gamma_5)$.  We denote left-handed ($L$) by
`$-$' and right-handed ($R$) by `$+$' for notational convenience.  The
complex couplings $a_{\pm}$ are expressed in terms of the
$\stau_{L}$-$\tau$-$\nef$ and $\stau_{R}$-$\tau$-$\nef$ couplings,
$a^{L}_{\pm}$ and $a^{R}_{\pm}$, respectively, as
\begin{equation}
 a_{\pm}= \cos\theta_{\stau}\cdot a^L_{\pm}
         +\sin\theta_{\stau}\cdot a^R_{\pm} \; .
\end{equation}
Here,
\begin{align}
\label{a_lr}
 & a_-^L=\tfrac{g}{\sqrt{2}}(U_{21}+U_{11}\tan\theta_W) \, ,
 & a_-^R=U_{31}\,Y_{\tau} \, , \no \\
 & a_+^R=-\tfrac{g}{\sqrt{2}}\,2\,U_{11}^*\tan\theta_W \, ,  
 & a_+^L=U_{31}^*\,Y_{\tau} \, ,
\end{align}
where $Y_{\tau}=-gm_{\tau}/(\sqrt{2}m_W\cos\beta)$ and $g$ is the
$SU(2)_L$ gauge coupling.  The couplings $a^L_-$ and $a^R_+$ come
from the stau-gaugino interactions, where chirality is conserved.  On
the other hand, $a^R_-$ and $a^L_+$ are proportional to the tau mass,
and come from the $\stau$-$\tau$-$\tilde H^0_d$ Yukawa coupling, which
flips chirality.  For further details see Ref.~\cite{Nojiri:1996fp}.

At this point it is worth noting that {\it either} the left-right stau
mixing {\it or} the gaugino-higgsino mixing of the neutralinos is
needed for the Majorana interference effects to be significant.  This
may be explained as follows: if both mixings are absent, the tau
produced in decay is either left-handed (gaugino interaction) or
right-handed (higgsino interaction), and the $\tau^+$ is of the
opposite chirality.  The two neutralinos then have opposite chirality
and their interference vanishes in the massless limit, where helicity
coincides with chirality.


\section{amplitudes and kinematics}
\label{sechel}

\begin{figure}[b]
\includegraphics[width=8cm,clip]{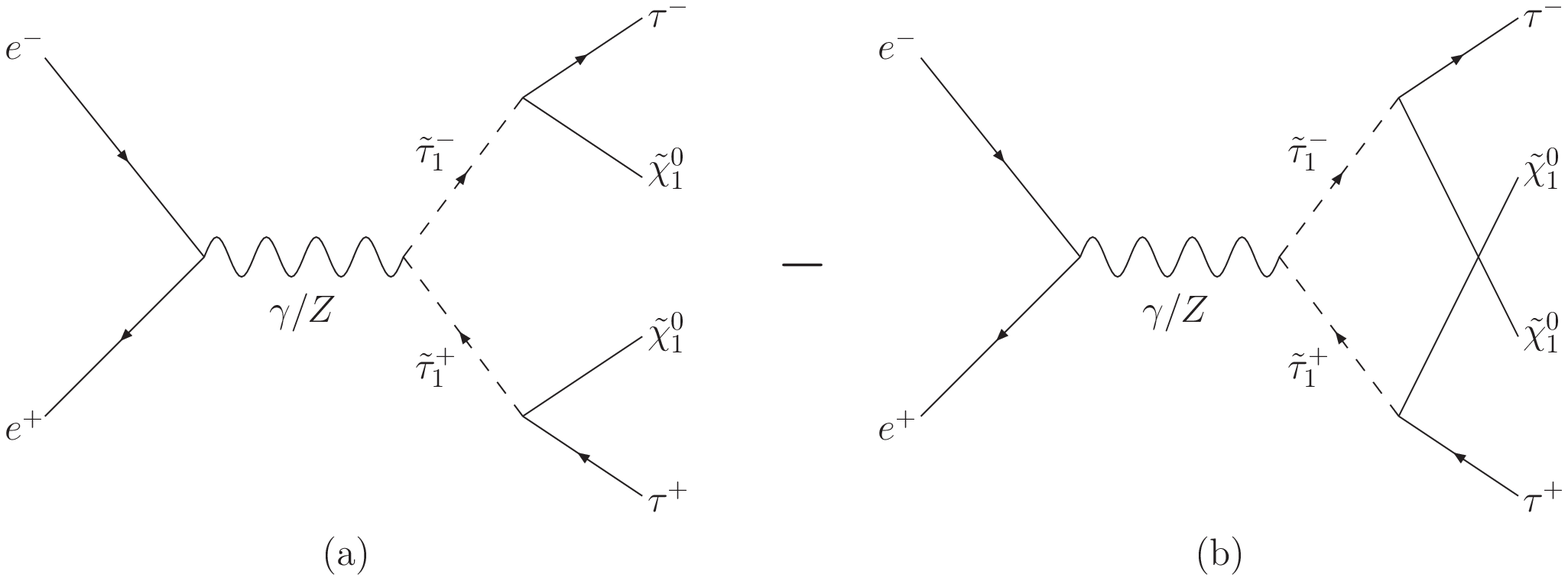}
\caption{\label{dres} Feynman diagrams for the process 
$e^+e^-\to\stau^+_1\stau^-_1\to\tau^+\tau^-\nef\nef$.}
\end{figure}

In this section we present helicity amplitude formulae for the process
\begin{align}
 & e^-(k,\lam) + e^+(\bar k,\bar\lam) \no \\
 & \to \stau_1^- + \stau_1^+ \no \\
 & \to \tau^-(k_1,\lam_1) + \tau^+(k_2,\lam_2) 
    +\nef(p_1,\sig_1) + \nef(p_2,\sig_2) \, ,
\end{align}
where the four-momentum and helicity of each particle are defined in
the center-of-mass (CM) frame of the $e^+e^-$ collision.  If
neutralinos are Majorana fermions, the two neutralinos ($\nef$) in the
final state are identical.  Therefore the crossed diagram (b) of
Fig.~\ref{dres} should be added to diagram (a) before the amplitude is
squared.  The relative sign between these two diagrams appears due to
Fermi statistics.

\begin{figure}[h]
 \includegraphics[width=8.5cm,clip]{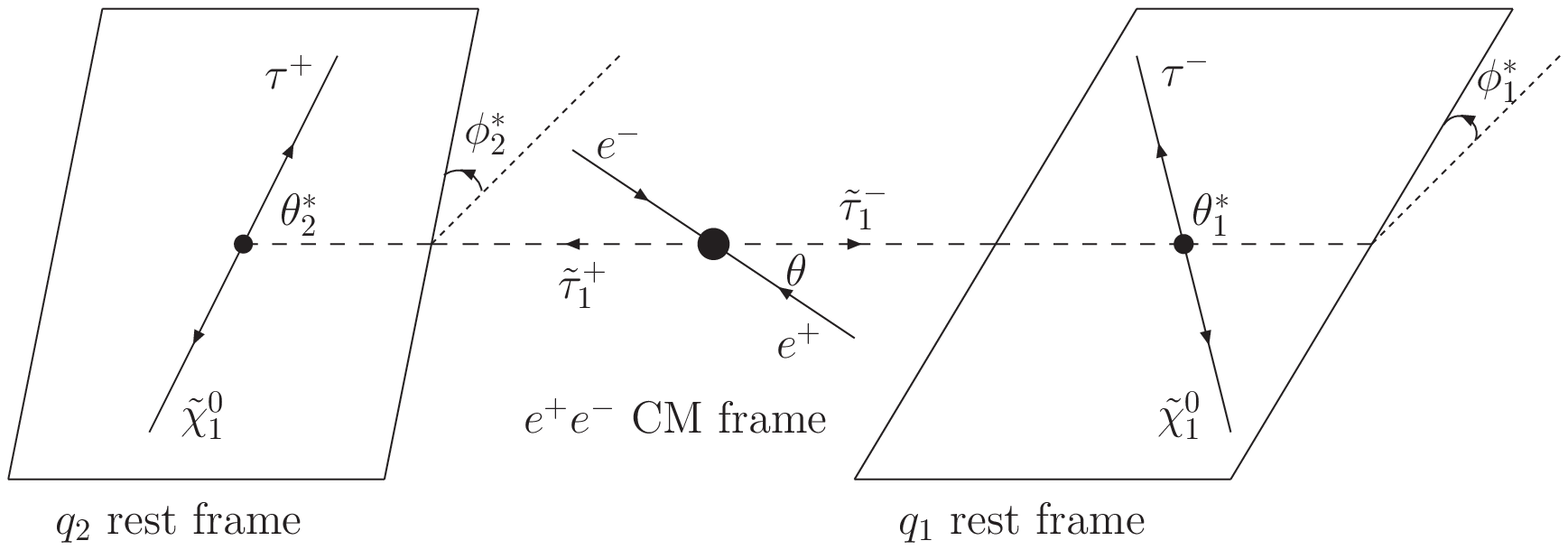}
 \caption{Schematic view of the coordinate system.}
\label{frame}
\end{figure}

\begin{widetext}

The full amplitude can be expressed as the product of the stau pair
production amplitude ($\cm_{\stau}$), the two Breit-Wigner propagators
for the staus,
\begin{equation}
 D_{\stau}(q^2)=(q^2-m_{\stau_1}^2+im_{\stau_1}\Gstau)^{-1} \, ,
\end{equation}
and two $\stau_1\to\tau\nef$ decay amplitudes ($\cm_{\neu}$ and
$\overline{\cm}_{\neu}$).  That is,
\begin{align}
  \cm
=&\cm (\lam;\lam_1,\lam_2,\sig_1,\sig_2) \no\\
=&\phantom{-l}\cm_{\stau}(\lam;q_1-q_2)\,
 D_{\stau}(q_1^2\phantom{'})D_{\stau}(q_2^2\phantom{'})\,
 \cm_{\neu}(q_1\phantom{'};\lam_1,\sig_1)\,
 \overline{\cm}_{\neu} (q_2\phantom{'};\lam_2,\sig_2) \no\\
&-\cm_{\stau}(\lam;q'_1-q'_2)\,
 D_{\stau}({q'_1}^2)D_{\stau}({q'_2}^2)\,
 \cm_{\neu}(q'_1;\lam_1,\sig_2)\,
 \overline{\cm}_{\neu} (q'_2;\lam_2,\sig_1) \no\\
=&\cm_1 - \cm_2 \, .
\label{amp}
\end{align}
The intermediate stau momenta can be written in terms of the 
final-state particle momenta: $q_1=k_1+p_1$, $q_2=k_2+p_2$,
 $q'_1=k_1+p_2$, 
and $q'_2=k_2+p_1$.  The stau pair production amplitude is given by
\begin{align}
  \cm_{\stau}(\lam;q)
  =\frac{e^2}{s}\Bigl[1+\frac{s}{s-m_Z^2+im_Z\Gamma_Z} 
   \,g_{\lam}\, (g_- \cos^2\theta_{\stau}+g_+\sin^2\theta_{\stau})\Bigr]\,
   \overline v(\bar k,-\lam)\!\not\!q\,u(k,\lam) \, ,
\label{amppro}
\end{align}
with $Z$ boson couplings to left- and right-handed charged leptons,
$g_-=\frac{-1+2s^2_W}{2s_Wc_W}$ and $g_+=\frac{s_W}{c_W}$,
respectively.  Here, we neglect the electron mass and take
$\bar\lam=-\lam$.  Using the straightforward Feynman rules for
Majorana fermions given in Ref.~\cite{Denner:1992vz}, the stau
decay amplitudes for $\cm_1$ are written as
\begin{align}
 & \cm_{\neu}(q_1;\lam_1,\sig_1)
  =\overline u(k_1,\lam_1)\,(a_-^*P_++a_+^*P_-)\,v(p_1,\sig_1) \, ,
\no \\
 & \overline{\cm}_{\neu} (q_2;\lam_2,\sig_2)
  =\overline u(p_2,\sig_2)\,(a_-P_-+a_+P_+)\,v(k_2,\lam_2) \, .
\label{decay1} 
\end{align}
Similarly, for $\cm_2$
\begin{align}
 & \cm_{\neu}(q'_1;\lam_1,\sig_2)
  =\overline u(k_1,\lam_1)\,(a_-^*P_++a_+^*P_-)\,v(p_2,\sig_2) \, ,
\no \\
 & \overline{\cm}_{\neu} (q'_2;\lam_2,\sig_1)
  =\overline u(p_1,\sig_1)\,(a_-P_-+a_+P_+)\,v(k_2,\lam_2) \, .
\label{decay2} 
\end{align}

In this article, we assume the lighter stau ($\stau_1$) to be the
next-to-lightest supersymmetric particle (NLSP), a common occurrence in
 many MSSM 
scenarios.  Hence, all $\stau_1$'s decay into $\tau$ plus $\nef$, and
the total decay width $\Gstau$ is just the partial width
$\Gamma(\stau_1\to\tau\nef)$.  Using the $\stau_1$ decay amplitude
$\cm_{\neu}$ in Eq.~(\ref{decay1}), the decay width is given by
\begin{align}
 \Gstau = \Gamma(\stau_1\to\tau\nef)  
 = \frac{1}{2m_{\stau_1}}\int\sum_{\lam_1,\sig_1}
       |\cm_{\neu}(q_1;\lam_1,\sig_1)|^2\,d\Phi_2  
 = \frac{1}{16\pi}\,\bigl(|a_-|^2+|a_+|^2\bigr)\,m_{\stau_1}
       \Bigl(1-\frac{m_{\nef}^2}{m_{\stau_1}^2}\Bigr)^2 \, ,
\label{width}
\end{align}
with $m_{\tau}=0$ and the Lorentz-invariant phase space factor
\begin{equation}
  d\Phi_n\Big(p=\sum_{i=1}^n k_i\Big)
 =(2\pi)^4\,\delta^4\Big(p-\sum_{i=1}^n k_i\Big)\,
  \prod_{i=1}^n\frac{d^3k_i}{(2\pi)^32k_i^0} \; .
\end{equation}
Throughout our study we neglect the tau mass, except in $Y_{\tau}$ of
the $\stau_1$-$\tau$-$\nef$ coupling, given in Eq.~(\ref{a_lr}).

Let us now define the kinematical variables.  In the CM frame of the
$e^+e^-$ annihilation, we choose the $\stau_1$ momentum direction as
the $z$-axis,
\begin{align}
 q_1 &= \tfrac{\sqrt{s}}{2}\,
      \bigl(1+\tfrac{q_1^2-q_2^2}{s},0,0,\beta\bigr) \, , \no\\
 q_2 &= \tfrac{\sqrt{s}}{2}\,
      \bigl(1+\tfrac{q_2^2-q_1^2}{s},0,0,-\beta\bigr) \, , 
\label{kine1}  
\end{align}
where $\beta=\bar\beta(\tfrac{q_1^2}{s},\tfrac{q_2^2}{s})$ with
$\bar\beta(a,b)\equiv(1+a^2+b^2-2a-2b-2ab)^{1/2}$, and we choose the
$\vec k\times\vec q_1$ direction as the $y$-axis.  For computational
convenience, we parametrize the momenta of $\tau^-$ and $\nef$ with
$p_1$ in the rest frame of $q_1$,
\begin{align}
 k_1^* &= \tfrac{\sqrt{q_1^2}}{2}\,
  \bigl(1-\tfrac{m_{\neu}^2}{q_1^2}, 
        \beta^*_1\sin\theta_1^*\cos\phi_1^*,
        \beta^*_1\sin\theta_1^*\sin\phi_1^*, 
        \beta^*_1\cos\theta_1^*\bigr) \, , \no\\
 p_1^* &= \tfrac{\sqrt{q_1^2}}{2}\,
  \bigl(1+\tfrac{m_{\neu}^2}{q_1^2},
        -\beta^*_1\sin\theta_1^*\cos\phi_1^*,
        -\beta^*_1\sin\theta_1^*\sin\phi_1^*,
        -\beta^*_1\cos\theta_1^*\bigr) \, ,  
\label{kine2}
\end{align}
with $\beta^*_1=1-m_{\nef}^2/q_1^2$.  Those of $\tau^+$ and $\nef$ with
$p_2$ are then in the $q_2$ rest frame,
\begin{align}
 k_2^*&= \tfrac{\sqrt{q_2^2}}{2}\,
  \bigl(1-\tfrac{m_{\neu}^2}{q_2^2},
       \beta^*_2\sin\theta_2^*\cos\phi_2^*,
       \beta^*_2\sin\theta_2^*\sin\phi_2^*,
       \beta^*_2\cos\theta_2^*\bigr) \, ,
\no\\ 
 p_2^*&= \tfrac{\sqrt{q_2^2}}{2}\,
  \bigl(1+\tfrac{m_{\neu}^2}{q_2^2},
       -\beta^*_2\sin\theta_2^*\cos\phi_2^*,
       -\beta^*_2\sin\theta_2^*\sin\phi_2^*,
       -\beta^*_2\cos\theta_2^*\bigr) \, ,
\label{kine3}
\end{align}
\end{widetext}
with $\beta^*_2=1-m_{\nef}^2/q_2^2$.  The two frames differ only by a
boost along the $z$-axis (see Fig.~\ref{frame}).  All the
frame-dependent variables with a star superscript ($*$) are those in
the $\stau_1$ rest frame.

Before turning to the numerical study, it is worthwhile to mention the
relation of the helicities between the taus ($\lam_{1,2}$) and the
neutralinos ($\sig_{1,2}$).  Generally, when a scalar (spinless)
particle decays into two fermions, they always have the same helicity
in the rest frame of the parent, due to helicity conservation in gauge
interactions.  In the massless neutralino limit, this relation remains
even in the $e^+e^-$ CM frame since the spin cannot flip due to the
boost, i.e.,
\begin{align}
 &\lam_1=\sig_1\ {\rm and}\ \lam_2=\sig_2\quad {\rm for}\ \cm_1 \, ,
\no\\ 
 &\lam_1=\sig_2\ {\rm and}\ \lam_2=\sig_1\quad {\rm for}\ \cm_2 \, .
\label{hrelation} 
\end{align}
Therefore, in the massless neutralino and tau limit, when $\tau^-$ and
$\tau^+$ have the same helicity ($\lam_1=\lam_2$), the two neutralinos
also have the same helicity, and we expect significant interference
effects between the two amplitudes, $\cm_1$ and $\cm_2$.  On the other
hand, there is no interference for the $\lam_1\ne\lam_2$ case since
either the amplitude $\cm_1$ or $\cm_2$ vanishes.  For finite
neutralino mass, however, the simple relations between the helicities
in Eq.~(\ref{hrelation}) do not hold due to the spin-flip effects
by Lorentz boosts.  Interference effects can then appear even for the
$\lam_1\ne\lam_2$ case.  In the following, we always sum over
neutralino helicities, but not tau helicities.  Notice that the tau
polarizations are observable statistically from their decay
distributions.

In order to confirm the above kinematical analysis, we consider the
interference term analytically.  In the spin-summed squared amplitude
$\sum|\cm|^2=\sum|\cm_1-\cm_2|^2$ (cf. Eq.~(\ref{amp})), the
interference term is given by the real part of $\cm_1\cm_2^*$ as
\begin{align}
 I=-2\ {\rm Re}\sum_{\lam,\sig_1,\sig_2}\cm_1\cm_2^* \, ,
\label{ifterm}
\end{align}
where we sum over the initial electron polarizations ($\lam$) and the
neutralino polarizations ($\sig_1$ and $\sig_2$), but keep the
$\tau^{\mp}$ helicities ($\lam_1$ and $\lam_2$) fixed.  The minus sign
is the result of Fermi statistics.  The interference term $I$ can be
expressed as the product of the production part $I_P$, the
Breit-Wigner propagator part, and the decay part $I_D$:
\begin{align}
 I=2\,{\rm Re}\,\big( I_P\cdot D_{\stau}(q_1^2)D_{\stau}(q_2^2)
   D_{\stau}^*({q'_1}^2)D_{\stau}^*({q'_2}^2)\cdot I_D \big) \, ,
\end{align}
where
\begin{align}
 I_P =& -\sum_{\lam} \cm_{\stau}(\lam;q_1-q_2)\,
                      \cm_{\stau}^*(\lam;q'_1-q'_2) \, ,
\label{ifp}\\
 I_D =&
 \sum_{\sig_1,\sig_2}
 \cm_{\neu}(q_1;\lam_1,\sig_1)\,
 \overline{\cm}_{\neu} (q_2;\lam_2,\sig_2) 
\no\\[-2.5mm]
 &\qquad\times
 \cm_{\neu}^*(q'_1;\lam_1,\sig_2)\,
 \overline{\cm}_{\neu}^* (q'_2;\lam_2,\sig_1) \, .
\end{align}
Note that we include the minus sign explicitly in the production part.
For the decay part, using the $\stau_1$ decay amplitudes of
Eqs.~(\ref{decay1}) and (\ref{decay2}), the relations
\begin{align}
 &\sum_s \bar u^T(p,s)\,\bar v(p,s)=C^{-1}(\!\not\!p-m) \, ,
\no\\
 &\sum_s v(p,s)\,u^T(p,s)=(\!\not\!p-m)\,C^T \, ,
\end{align}
and the properties of the charge conjugation matrix $C$, we obtain 
\begin{align}
 I_D=
 \begin{cases}
  |a_-|^2|a_+|^2\ 
  {\rm tr}\big[\!\not\!k_1 \!\not\!p_1 \!\not\!k_2 \!\not\!p_2 \big] 
   & {\rm for}\ \lam_1=\lam_2 \, ,\\
  \tfrac{1}{2}\, m_{\nef}^2\,\bigl(|a_-|^4+|a_+|^4\bigr)\ 
  {\rm tr}\big[\!\not\!k_1 \!\not\!k_2\big] 
   & {\rm for}\ \lam_1\ne\lam_2 \, .
 \end{cases} 
\label{ifd}
\end{align}
This confirms the above kinematical analysis.


\section{Total cross sections}
\label{sectot}

\begin{figure}
\includegraphics[width=8cm,clip]{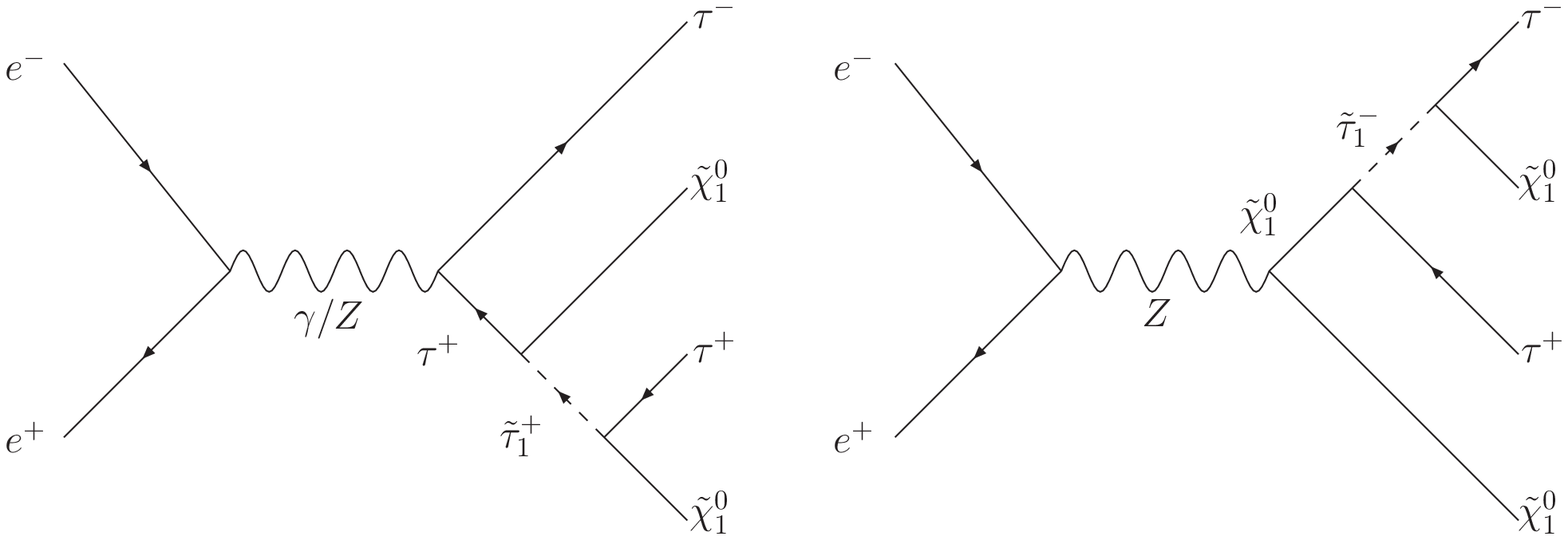}
\caption{The single-resonance contributions to the final state 
$\tau^+\tau^-\nef\nef$ in $e^+e^-$ annihilation.}
\label{sres}
\end{figure}

In this section, we present total cross section for the process
$e^+e^-\to\tau^+\tau^-\nef\nef$ around and well below the $\stau_1$
pair production threshold.  In addition to the double-resonance
contribution (Fig.~\ref{dres}) which we have discussed so far, we
further consider the single-resonance contributions to the final
states, shown in Fig.~\ref{sres}.  There are two additional diagrams,
each having its own crossed diagram for the Majorana neutralino case.

The cross section for the process $e^+e^-\to\tau^+\tau^-\nef\nef$
averaged over initial electron polarization and summed over neutralino
polarizations reads
\begin{align}
 d\sig^{\lam_1\lam_2}=\frac{1}{2s}\frac{1}{4}
 \sum_{\lam,\sig_1,\sig_2}
 |\cm(\lam;\lam_1,\lam_2,\sig_1,\sig_2)|^2\,\frac{1}{2}\,d\Phi_4 \, ,
\label{totalx}
\end{align}
where $\lam_1$ and $\lam_2$ are the helicities of $\tau^-$ and
$\tau^+$.  In addition to the initial-state spin-average factor, we
divide by the statistical factor for two identical neutralinos in the
final state.  The amplitude $\cm$ is summed over all diagrams; not
only the double-resonance contributions in Eq.~(\ref{amp}), but also
the single-resonance contributions.  The four-body phase space factor
$d\Phi_4$ can be decomposed as the two-body phase space $d\Phi_2$, and
can be parametrized by the kinematical variables defined in
Eqs.~(\ref{kine1})-(\ref{kine3}) as
\begin{align}
 d\Phi_4=&d\Phi_2(k+\bar k=q_1+q_2)\,
               \frac{dq_1^2}{2\pi}\frac{dq_2^2}{2\pi}\no\\
       &\times d\Phi_2(q_1=k_1+p_1)\, d\Phi_2(q_2=k_2+p_2) \no\\
 =&\frac{\beta\,\beta^*_1\,\beta^*_2}{2\pi(32\pi^2)^3}\, dq_1^2dq_2^2\,
 d\cos\theta\,d\cos\theta^*_1 d\phi_1^*\,d\cos\theta^*_2 d\phi_2^* \, ,
\end{align}
where $\theta$ is the scattering angle between the $e^-$ momentum
($\vec k$) and $\vec q_1$ in the $e^+e^-$ CM frame (see
Fig.~\ref{frame}).

Now let us explain several parameters we use in our numerical
analysis.  All the helicity amplitudes, including the single-resonance
contributions, are calculated by HELAS subroutines~\cite{HELAS}, and
numerical integrations are done with the help of the Monte Carlo
integration package BASES~\cite{Kawabata:1995th}. We
fix the $\stau_1$ mass at $m_{\stau_1}=150$~GeV.  As for the SUSY
parameters, including the left-right stau mixing, we take the
following values so that the Majorana effects are expected to be
large:
\begin{align}
 \theta_{\stau}=45^{\circ} \, ,\ \tan\beta=50 \, ,\
  M_2=300\ {\rm GeV} \, ,\ \mu=70\ {\rm GeV} \, ,
\label{SUSYpara}
\end{align} 
and adopting the mass relation $M_1=\frac{5}{3}\tan^2\theta_WM_2$.
This parameter set corresponds to a higgsino-like neutralino LSP.  For
these parameters, the $\nef$ mass is $m_{\nef}=50$~GeV, which yields a
$\stau_1$ total decay width of $\Gstau=0.52$~GeV, using
Eq.~(\ref{width}).

\begin{figure}
\includegraphics[width=8.5cm,clip]{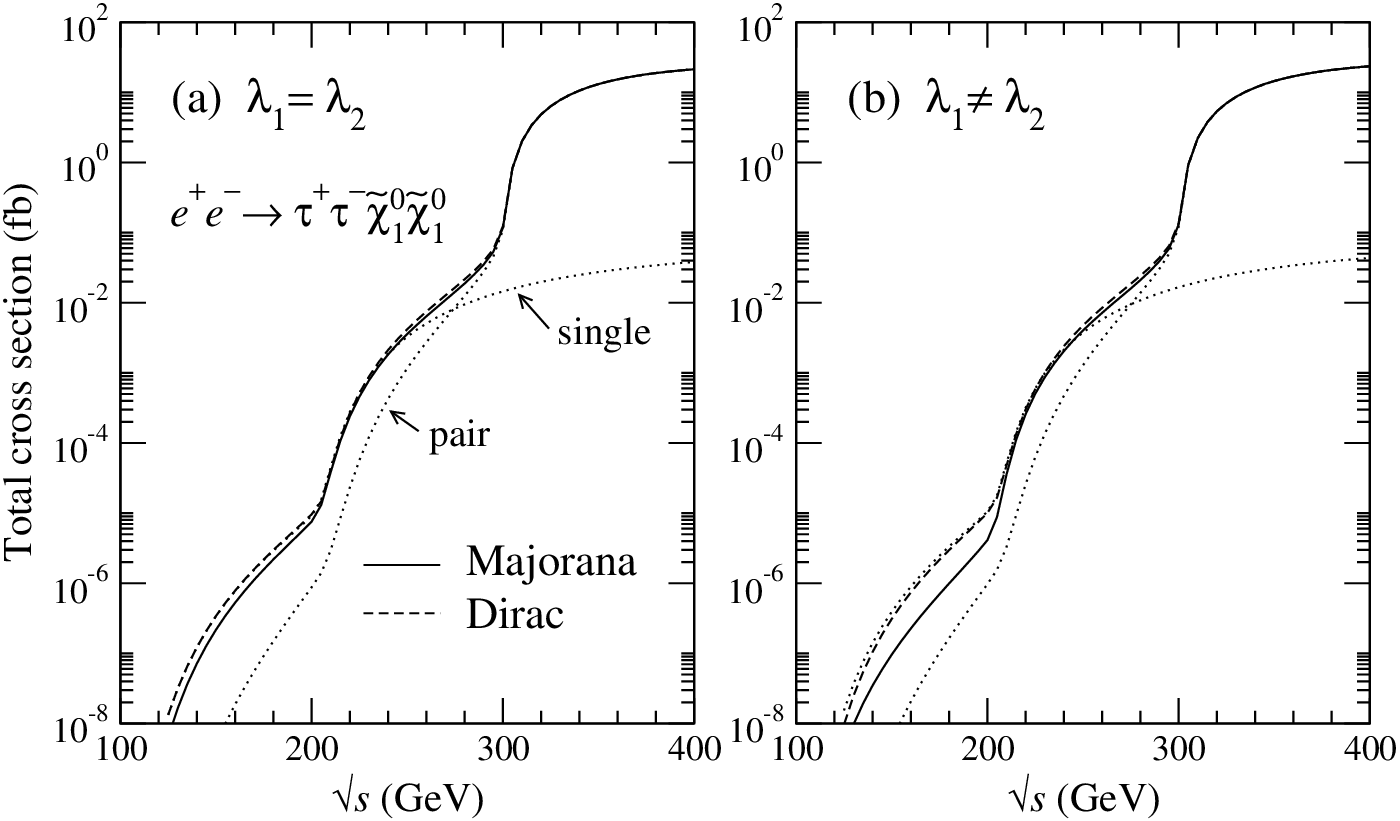}
\caption{Total cross section as a function of collision energy, of 
the process $e^+e^-\to\tau^+\tau^-\nef\nef$, for (a) $\lam_1=\lam_2$
and (b) $\lam_1\ne\lam_2$, where $\lam_{1,2}$ is the helicity of the
$\tau^{\mp}$.  Solid and dashed lines show the case that neutralinos
are Majorana and Dirac fermions, respectively.  Dotted lines are for
each contribution from the $\stau_1$ pair and singly resonant
$\stau_1$ production without the interference term.}
\label{total}
\end{figure}

Figure \ref{total} shows the $\tau^{\pm}$ helicity-dependent total cross
section, Eq.~(\ref{totalx}), for the process
$e^+e^-\to\tau^+\tau^-\nef\nef$ as a function of the CM energy
($\sqrt{s}$) in $e^+e^-$ collisions.  In order to show the
single-resonance contributions, the cross section is shown starting
from rather low values.  Solid lines denote the cross sections
including the crossed diagrams, such as Fig.~\ref{dres}(b), which
should be present for Majorana neutralinos.  The dashed lines are
obtained by neglecting the crossed diagrams, which corresponds to
Dirac neutralinos.  As a reference, each contribution from the
$\stau_1$ pair and the single-$\stau_1$ production
(without the interference term) is shown by dotted lines.

Above the $\stau_1$ pair production threshold
($\sqrt{s}>2m_{\stau_1}=300$ GeV) the contribution from $\stau_1$ pair
production (Fig.~\ref{dres}) is dominant.  Below pair threshold the
single-resonance processes (Fig.~\ref{sres}) contribute dominantly,
even though the cross section becomes very small.  The reduction of
the cross section by the interference effects for Majorana neutralinos
can be seen below the $\stau_1$ pair and the single $\stau_1$
production thresholds.  Unfortunately, the total cross section in the
region below threshold where Majorana interference effects become
important is not at a level which could be observed.


\section{Majorana effects}
\label{secdif}

\begin{figure*}
\includegraphics[width=16cm,clip]{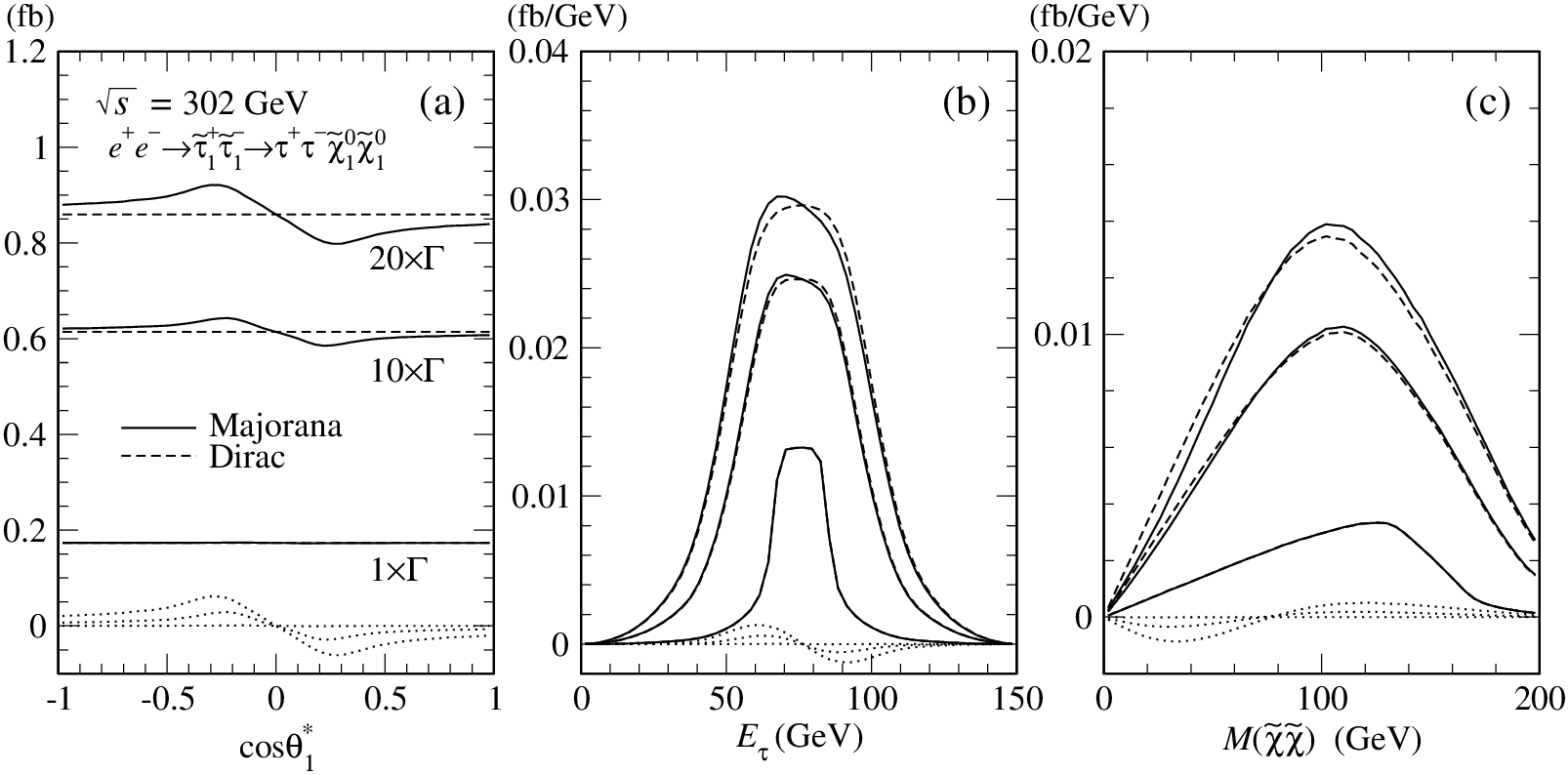}
\caption{Distributions of (a) $\cos\theta_1^*$, (b) $E_{\tau}$, and 
(c) $\im$ for the process
$e^+e^-\to\stau_1^+\stau_1^-\to\tau^+\tau^-\nef\nef$ at
$\sqrt{s}=302$~GeV for $1,10,20\times\Gstau$ in the massless neutralino
limit.  Solid and dashed lines show the Majorana and Dirac
neutralino cases, respectively.  Also shown is the behavior of the
interference term by dotted lines.}
\label{staucei}
\end{figure*}

We have seen that it is difficult to observe a Majorana interference
effect in the total cross section.  In this section, we therefore
study in detail the kinematical correlations due to interference
effects that appear only for Majorana neutralinos.  We present several
distributions for the process $e^+e^-\to\tau^+\tau^-\nef\nef$ near the
stau pair production threshold, and discuss the Majorana effects as a
function of the finite $\stau_1$ width.  We consider the following
three distributions to see the interference:
\renewcommand{\labelenumi}{(\alph{enumi})} 
\begin{enumerate}
 \item $\cos\theta_1^*$: defined in the $q_1$ rest frame
       in Eq.~(\ref{kine2}) (See Fig.~\ref{frame});
 \item $E_{\tau}$: the $\tau^{\pm}$ energy in the
       laboratory frame;
 \item $\im$: the invariant mass of the neutralino pair,
       $M^2=(p_1+p_2)^2=(k+\bar k-k_1-k_2)^2 \,$.
\end{enumerate}
Note that, from the experimental point of view, these variables are
not observables, since taus always decay into at least one neutrino,
which escapes detection together with the neutralinos; the
kinematical system is unconstrained by observables and cannot be
reconstructed.  Therefore, our studies may be regarded as pedagogical,
or may apply to models where the neutralino momenta can be
kinematically reconstructed, such as those with $R$-parity violation.

To begin with, for simplicity we consider only the $\stau_1$ pair
production process
$e^+e^-\to\stau_1^+\stau_1^-\to\tau^+\tau^-\nef\nef$, and take the
massless neutralino limit.

Figure \ref{staucei} shows the distributions of (a) $\cos\theta_1^*$,
(b) $E_{\tau}$, and (c) $\im$, at $\sqrt{s}=302$ GeV for the
$\lam_1=\lam_2$ case.  We use the same SUSY parameter set in
Eq.~(\ref{SUSYpara}) as the total cross sections.  To examine the
finite $\stau_1$ width effect, we vary the total width
$\Gstau$ as $1,10,20\times\Gstau$.  In the limit of
$m_{\nef}=0$, the $\stau_1$ total decay width is
$\Gstau=0.66$~GeV from Eq.~(\ref{width}), hence
$10\times\Gstau=6.6$~GeV and $20\times\Gstau=13$~GeV.
Above threshold, the wider the decay width, the larger the cross
section.  Solid and dashed lines show the Majorana and Dirac
neutralino cases, respectively.  Also shown as a reference is the
behavior of the interference term only, Eq.~(\ref{ifterm}), by dotted
lines.  For the
realistic $1\times\Gstau$ case, the interference patterns are
barely discernible in Fig.~\ref{staucei}.  The contribution of the
interference term to the cross section is at most about $0.1\%$, $3\%$
and $6\%$ for $1,10{\rm\ and}\ 20\times\Gstau$, respectively.
The interference effect is roughly proportional to the $\stau_1$ decay
width.  The following features are worth noting: (i) For the
$\lam_1\ne\lam_2$ case, there is no interference, as expected from the
discussion of Sec.~\ref{sechel}.  (ii) For the Dirac neutralino case,
the cross section does not depend on $\cos\theta_{1}^*$ or
$\cos\theta_{2}^*$ since the two produced staus decay independently.
On the other hand, for the Majorana case, the distribution is no
longer flat.  The $\cos\theta_2^*$ distribution is the same as that of
$\cos\theta_1^*$, but for a relative sign, according to
$CP$-invariance.  (iii) A major portion of the interference pattern
disappears when we integrate out the kinematical variables, such as
$\cos\theta_1^*$, $E_\tau$, or $M(\neu\neu)$.  This is why one can
hardly view the correlation effect in the total cross section above
pair production threshold in Fig.~\ref{total}.  (iv) As for the
$\sqrt{s}$ dependence of the interference effect, it exists even at
higher energies.  However, the cross section also grows, and the
relative effect of the interference term becomes smaller.  Therefore
the Majorana effect can be seen only near threshold, even though the
cross section is small due to the $p$-wave threshold factor of
$\beta^3$.

Now we attempt to explain why the interference term changes sign as 
in Fig.~\ref{staucei}.  We take two different approaches.  One
is to consider the interference term, Eq.~(\ref{ifterm}),
analytically.  Another is, more physically, to investigate them
kinematically.

First, let us consider the $\stau_1$ pair production part of the
interference term, Eq.~(\ref{ifp}), analytically.  Using the
$\stau_1$ pair production amplitude of Eq.~(\ref{amppro}),
\begin{align} 
 I_P=& -\frac{e^4}{s^2}\sum_{\lam}\,[\cdots]^2\,
 \overline v(\bar k,-\lam)\,(\!\not\!q_1-\!\not\!q_2)\,u(k,\lam)
\no\\[-0.25cm]
 &\qquad\qquad\qquad\times 
  \overline u(k,\lam)\,(\!\not\!q'_1-\!\not\!q'_2)\,
           v(\bar k,-\lam)
\no\\
 =&-\frac{4e^4}{s^2}[\cdots]^2\ {\rm tr}
   \big[\!\not\!q_1 \!\not\!k \!\not\!q'_1 \!\not\!\bar k\,\big] \, .
\end{align} 
The trace part is given by the dot-products of the kinematical
variables in Eqs.~(\ref{kine1})-(\ref{kine3}).  The dependence on the
scattering angle $\theta$ appears only in this part.  We integrate it
out and obtain
\begin{align}
 \int_{-1}^{1} I_P\, d\cos\theta \qquad &
\no \\
 = -\frac{4\beta e^4}{3} [\cdots]^2
 \Bigl[&+(1-\tfrac{m_{\neu}^2}{q_1^2})
            (1+\tfrac{q_1^2-q_2^2}{s})\cos\theta_1^*
\no \\ 
  &-(1-\tfrac{m_{\neu}^2}{q_2^2})
            (1+\tfrac{q_2^2-q_1^2}{s})\cos\theta_2^*
\no \\
  &-m_{\neu}^2\beta\,(\tfrac{1}{q_1^2}+\tfrac{1}{q_2^2})\Bigr] \, ,
\label{ifp2}
\end{align}
where the neutralino mass is kept explicitly for later discussion.
This shows that the interference term behaves like $-\cos\theta_1^*$,
or $+\cos\theta_2^*$ in the massless neutralino limit.  This is
consistent with Fig.~\ref{staucei} (a), where one can see negative
interference for $\cos\theta_1^*>0$ and positive interference for
$\cos\theta_1^*<0$. 

Next, we turn to the kinematical approach to understand the Majorana
effects more physically.  The question is if there is a case when both
amplitudes $\cm_1$ and $\cm_2$ of Eq.~(\ref{amp}) are large, thus the
total amplitude $\cm=\cm_1-\cm_2$ is suppressed due to the relative
sign from Fermi statistics.  For instance, the amplitude $\cm$ should
be suppressed when the two neutralinos have identical spins and
four-momenta.  The same spin is realized by setting the $\tau^-$ and
$\tau^+$ helicities equal in the massless neutralino limit, as
previously noted.  Let us determine the condition for which two
neutralinos have the same four-momentum, i.e., $p_1=p_2$.  It is
obvious that the relative azimuthal angle is zero,
$\phi^*\equiv\phi_2^*-\phi_1^*=0^{\circ}$.  For simplicity, we
consider the limit that both staus are on mass shell,
$q_1^2=q_2^2=m_{\stau_1}^2$, where the amplitude $\cm_1$ is
significant.  In this limit, we can find the following simple
kinematical point:
\begin{align}
 \cos\theta_2^*=-\cos\theta_1^* \, , \quad
 \cos\theta_1^*=\beta,\quad \phi_2^*=\phi_1^* \, ,
\label{cosa}
\end{align}
with $\beta=(1-4m_{\stau_1}^2/s)^{1/2}$.  At this point, not only
$q_1^2$ and $q_2^2$ but also ${q'_1}^2$ and ${q'_2}^2$ are on
$\stau_1$ mass shell:
\begin{equation}
 q_1^2=q_2^2={q'_1}^2={q'_2}^2=m_{\stau_1}^2 \, .
\end{equation}
We note that the propagator momenta squared of the crossed diagram are
expressed as
\begin{multline}
 {q'_{1,2}}^2=\frac{s}{8}
 \Big[(1+\beta^2)(1+\cos\theta_1^*\cos\theta_2^*) \\
 +(1-\beta^2)\sin\theta_1^*\sin\theta_2^*\cos\phi^*\\
 \pm 2\beta(\cos\theta_1^*+\cos\theta_2^*)\Big] \, .
\label{q34}
\end{multline}
\begin{figure}
\includegraphics[width=8.5cm,clip]{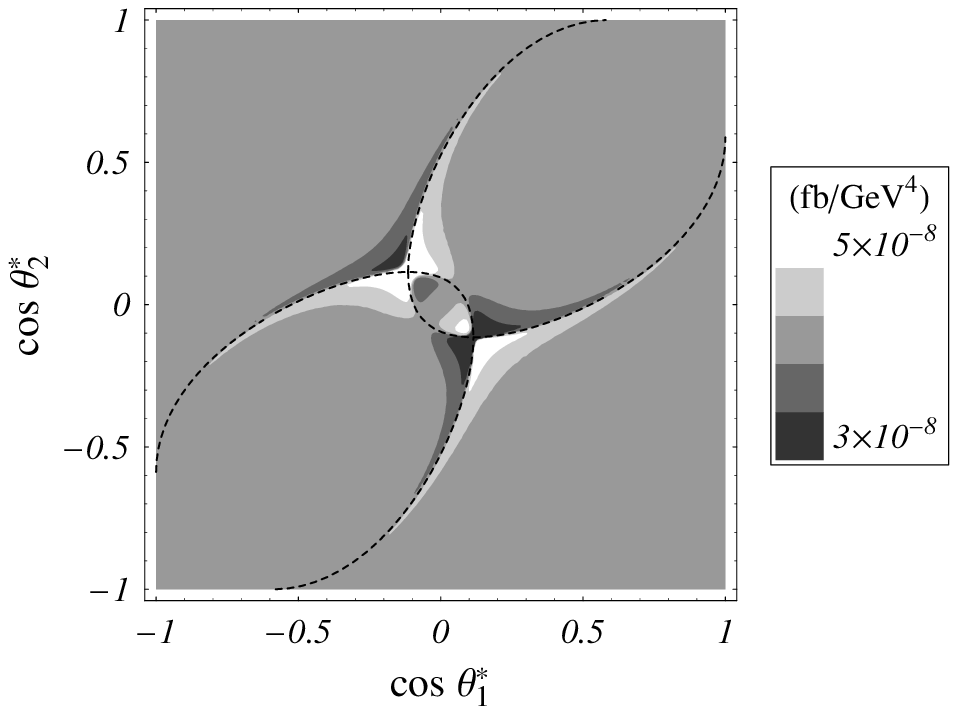}
\caption{Contour plot of the differential cross section, 
Eq.~(\ref{conteq}), for the process
$e^+e^-\to\stau_1^+\stau_1^-\to\tau^+\tau^-\nef\nef$ at
$\sqrt{s}=302$~GeV in the $m_{\nef}=0$ limit when
$q_1^2=q_2^2=m_{\stau_1}^2$ and $\phi^*=0^{\circ}$, is shown in the
$\cos\theta^*_1$-$\cos\theta^*_2$ plane, where $\cos\theta^*_{1,2}$
are the kinematical variables defined in the $q_{1,2}$ rest frame.
${q'_{1}}^2=m_{\stau_1}^2$ and ${q'_{2}}^2=m_{\stau_1}^2$ are also
shown by dashed lines.}
\label{contour}
\end{figure}

Figure \ref{contour} is a contour plot of the differential cross section
for the process $e^+e^-\to\stau_1^+\stau_1^-\to\tau^+\tau^-\nef\nef$
at $\sqrt{s}=302$~GeV in the massless neutralino limit when
$q_1^2=q_2^2=m_{\stau_1}^2$ and $\phi^*=0$,
\begin{align}
 \frac{d\sigma^{\lam_1\lam_2}}
 {dq_1^2dq_2^2\, d\cos\theta^*_1\, d\cos\theta^*_2d\phi^*}
 \bigg|_{q_1^2=q_2^2=m_{\stau_1}^2,\, \phi^*=0} \, ,
\label{conteq}
\end{align}
for the $\lam_1=\lam_2$ case. Using Eq.~(\ref{q34}), we also show the
${q'_{1}}^2=m_{\stau_1}^2$ and ${q'_{2}}^2=m_{\stau_1}^2$ trajectories
by dashed lines.  The interference effect can be seen along these
lines, especially around the intersection points, where both
${q'_1}^2$ and ${q'_2}^2$ approach $m_{\stau_1}^2$ and the effect is
largest.  This arises from the double Breit-Wigner factor
$D({q'_1}^2)D({q'_2}^2)$ of the crossed amplitude $\cm_2$.  Note also
that the sign of the interference term changes over the
${q'_{1}}^2=m_{\stau_1}^2$ or ${q'_{2}}^2=m_{\stau_1}^2$ trajectory
because of the Breit-Wigner resonant factor.  Around one of the
intersections (black region), which corresponds to the kinematical
point of Eq.~(\ref{cosa}) with $\beta=0.115$, the cross section is
strongly suppressed because the two neutralinos have the same momenta.
On the other hand, it is enhanced around another intersection (white
region), i.e., $(\cos\theta^*_1,\cos\theta^*_2)=(-\beta,\beta)$.  That
is because the four-momenta of the $\tau^-$ and $\tau^+$ leptons
become identical at this point, and the two interfering amplitudes are
constructive.  Here one can also see negative interference for
$\cos\theta_1^*>0$ ($\cos\theta_2^*<0$) and constructive interference
for $\cos\theta_1^*<0$ ($\cos\theta_2^*>0$).

\begin{figure*}
\includegraphics[width=16cm,clip]{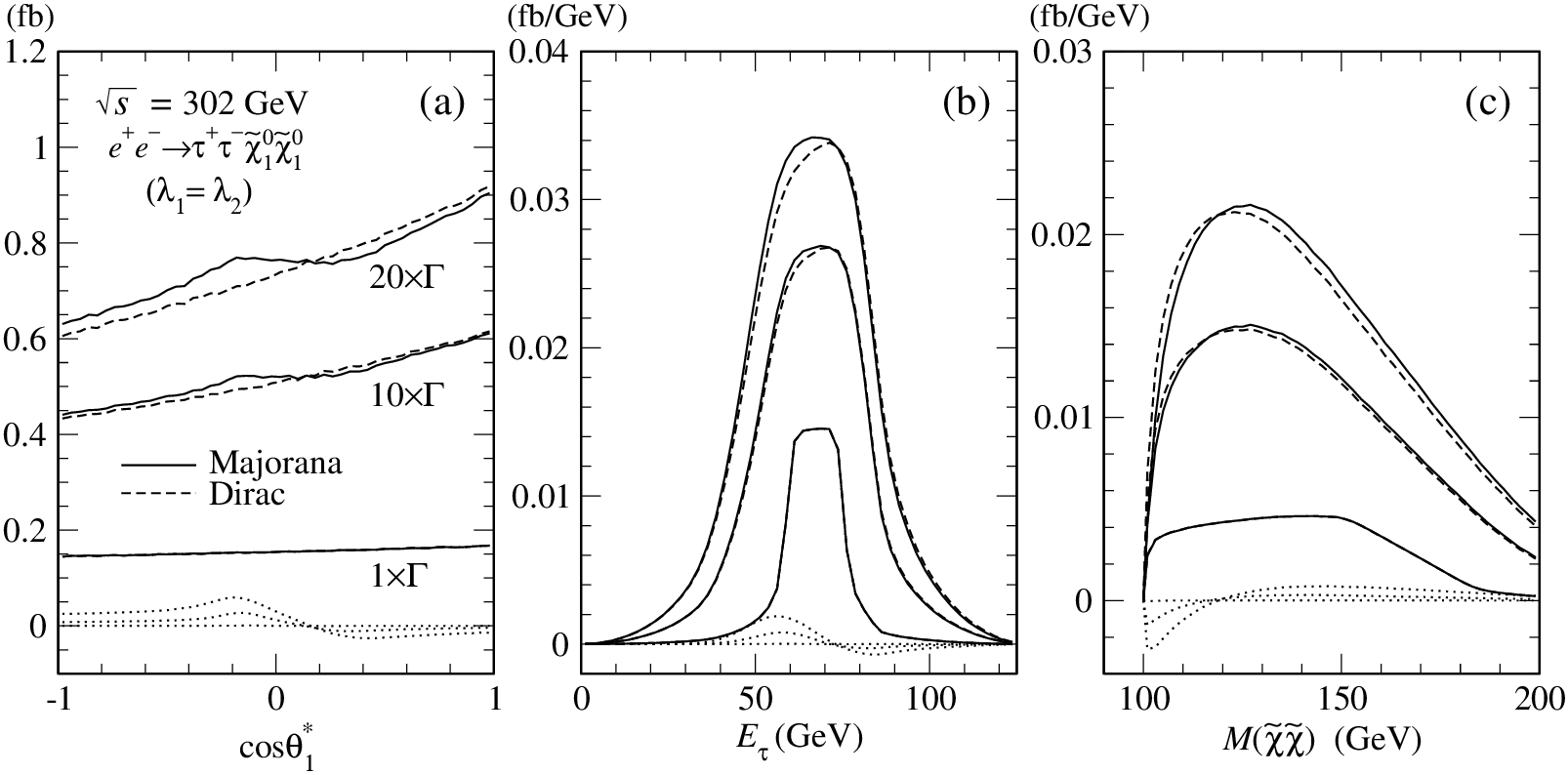}
\caption{Distributions of (a) $\cos\theta_1^*$, (b) $E_{\tau}$, and
(c) $\im$ for the process $e^+e^-\to\tau^+\tau^-\nef\nef$ for the
$\lam_1=\lam_2$ case at $\sqrt{s}=302$~GeV for
$1,10,20\times\Gstau$.  Solid and dashed lines show the
Majorana and Dirac neutralino cases, respectively. Also shown is the
behavior of the interference term by dotted lines.}
\label{allcei}
\end{figure*}

Let us now try to explain the consistency among the three
distributions $\cos\theta_1^*$, $E_{\tau}$ and $\im$ in
Fig~\ref{staucei}. In the phase space limit of $\im=2m_{\nef}=0$, the
four-momenta of the two neutralinos are identical ($p_1=p_2$).  As
discussed above, therefore, negative interference should be expected
near this limit due to Fermi statistics.  Furthermore, when
${q'_{1}}^2={q'_{2}}^2=m_{\stau_1}^2$ in this limit, it is obvious
that the amplitude is suppressed in the region $\cos\theta_1^*>0$ and
$\cos\theta_2^*<0$.  In addition, due to boost effects, this
kinematical region corresponds to large $E_{\tau}$.

So far all our distributions have been given for massless neutralinos.
Let us now show results for finite neutralino masses.

Figure \ref{allcei} is the same as Fig.~\ref{staucei}, except we
include the single-resonance contributions to the final state and a
finite neutralino mass, $m_{\nef}=50$ GeV.  Since the single-resonance
contributions are not so small just above the stau pair threshold (see
Fig.~\ref{total}), the distributions are affected significantly.  As
for the $\im$ distribution, the lowest value of the invariant mass is
$2m_{\nef}=100$ GeV in this case.  We find that the interference
pattern is basically the same as in Fig.~\ref{staucei}, where we
consider only the $\stau_1$ pair production process in the
$m_{\nef}=0$ limit.

However, one can also see a difference in the behavior of the
interference between Fig.~\ref{allcei} and Fig.~\ref{staucei}.  The
region of constructive interference becomes larger than that in
Fig.~\ref{staucei}.  The reason is given by Eq.~(\ref{ifp2}): the third
term proportional to the neutralino mass squared is
additive with the first term, $-\cos\theta_1^*$, when we consider a
finite neutralino mass.

We can also repeat the kinematical analysis for the Majorana case,
taking into account a finite neutralino mass. Equation (\ref{cosa}) is
slightly modified by the boost effect as
\begin{align}
 \cos\theta_2^*=-\cos\theta_1^* \, ,\quad
 \cos\theta_1^*=
    \frac{1+m_{\nef}^2/m_{\stau_1}^2}{1-m_{\nef}^2/m_{\stau_1}^2}\,\beta
  >\beta \, , 
\end{align}
where the two neutralinos have the same four-momentum.  On the other
hand, the condition that the momenta of $\tau^-$ and $\tau^+$ are the
same does not change, namely
$(\cos\theta^*_1,\cos\theta^*_2)=(-\beta,\beta)$.  This supports the
tendency of the interference pattern to increase with finite
neutralino mass in Fig.~\ref{allcei}.

\begin{figure*}
\includegraphics[width=16cm,clip]{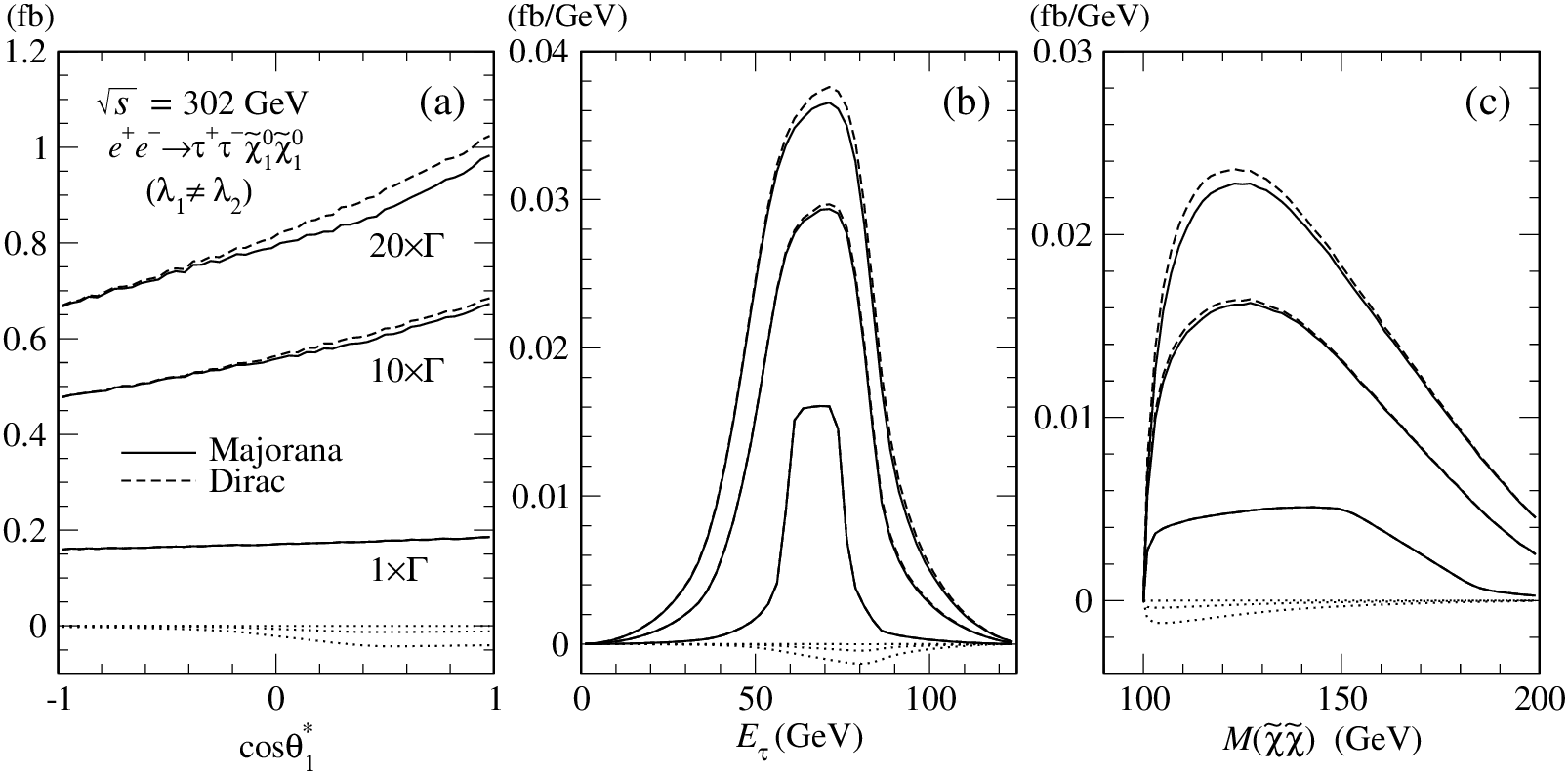}
\caption{The same as Fig.~\protect\ref{allcei}, but for the 
$\lam_1\ne\lam_2$ case.}
\label{all2cei}
\end{figure*}

Figure \ref{all2cei} is the same as Fig.~\ref{allcei}, but for the
$\lam_1\ne\lam_2$ case.  As we discussed and explicitly showed in
Eq.~(\ref{ifd}) at the end of Sec.~\ref{sechel}, one can see the
Majorana interference effect even for the $\lam_1\ne\lam_2$ case.  The
distributions for the $\lam_1=\lam_2$ case are much more useful than
those of the $\lam_1\ne\lam_2$ case because in (a) there is a
significant shape change which includes two inflection points, whereas
the distribution is monotonic in the latter case; and because in (b,c)
the distributions for $\lam_1=\lam_2$ involve a peak shift, not just a
magnitude change as for $\lam_1\ne\lam_2$.

A final question we may ask is, what would be the observability of the
Majorana interference effect?  We answer this by calculating the
integrated luminosity required at a future linear collider to observe
a $3\sigma$ effect, using the case MSSM described above and collisions
at $\sqrt{s}=302$~GeV.  The simplest observable is the $\cos\theta^*_1$
distribution of Fig.~\ref{staucei}(a), which has a small
forward-backward asymmetry $A$.
We furthermore make the estimate using the nonphysical
enhanced-effect case of $10\times\Gstau$.  The formula for the
statistical uncertainty on an asymmetry measurement is,
\begin{equation}
 \triangle A \; = \; \sqrt{ N_F(2N_B/N^2)^2 + N_B(2N_F/N^2)^2 } \; ,
\end{equation}
where $N_F(N_B)$ is the number of events with $\cos\theta^*_1>0$
($\cos\theta^*_1<0$) and $N=N_F+N_B$ is the total number of events.
Since the asymmetry is small, 0.02 in this case, we may use the
approximation $N_F=N_B=N/2$.  Substituting and rearranging, we
arrive at a formula for the required luminosity to observe a $3\sigma$
effect:
\begin{equation}
 L_{\rm min} \; = \; (3/A)^2/\sigma_{\rm total} \; .
\end{equation}
Plugging in $A=0.02$ and $\sigma_{\rm total}=1.2$~fb, we arrive at an
estimate of 
18,750~fb$^{-1}$.  One is swift to conclude that even in an
enhanced-effect scenario due to abnormally large finite stau width, a
future linear collider unfortunately cannot observe this effect.


\section{Summary and discussions}
\label{secsum}

We studied the quantum mechanical correlation between two identical
neutralinos, which exists only if they are Majorana particles, for the
process $e^+e^-\to\stau^+_1\stau^-_1\to\tau^+\tau^-\nef\nef$.  We also
considered the single-resonance contributions to the final state.

We found that the correlation between two neutralinos appears near
$\stau_1$ 
pair production threshold in the presence of a finite stau width and
mixing of the staus and/or neutralinos.  We discussed the finite width
effect in detail, and found that the correlation effect tends to be
proportional to the decay width.  Distributions in several kinematical
variables, such as $\cos\theta_1^*$, $E_{\tau}$, and $M(\neu\neu)$ as
defined in Sec.~\ref{secdif}, show the interference effect, although
the effect largely disappears after integrating over these
distributions.  Because the correlation effects are significant only
in a specific kinematical configuration, they can be observed only
in models where the neutralino momenta can be kinematically
reconstructed, such as in models with $R$-parity violation.
Unfortunately the interference pattern does not persist in the 
angular distribution of the final-state taus in the lab frame,
disallowing these observables to be determined in an
$R$-parity-conserving MSSM scenario.

Our brief estimate of the potential observability of the Majorana
interference effect (assuming an $R$-parity-violating scenario) using
the asymmetry of the stau decay angular distribution in its rest frame
is unfortunately not optimistic.  It appears that an unrealistic
amount of integrated luminosity at a future $e^+e^-$ collider
operating at stau pair threshold would be required.  Thus this
particular interference effect for stau NLSP pairs remains a
pedagogical observation, but a very interesting one nonetheless.

Before closing our discussions, we point out that a significant
interference effect is also expected in $\tilde e_L^{\pm}\tilde
e_R^{\mp}$ and $\tilde\mu_L^{\pm}\tilde\mu_R^{\mp}$ pair production
processes.  The quantitative study will be reported elsewhere.


\begin{acknowledgments}
K.M.\ would like to thank M.~Aoki, E.~Senaha, H.~Shimizu, and H.~Yokoya
for discussions and encouragement.  The work of K.H.\ is supported in part
by the Grant-in-Aid for Scientific Research, Ministry of Education,
Culture, Science and Technology, Japan (No 17540281).
\end{acknowledgments}




\newpage

\begin{thebibliography}{00}

\bibitem{Dawson:1983fw}
  S.~Dawson, E.~Eichten and C.~Quigg,
  Phys.\ Rev.\ D {\bf 31}, 1581 (1985).

\bibitem{Beenakker:1996ed}
  W.~Beenakker, R.~Hopker and M.~Spira,
  arXiv:hep-ph/9611232.
\bibitem{Beenakker:1999xh}
  W.~Beenakker {\it et al.}, 
  Phys.\ Rev.\ Lett.\  {\bf 83}, 3780 (1999);
  Nucl.\ Phys.\ B {\bf 515}, 3 (1998).

\bibitem{Hinchliffe:1996iu}
  I.~Hinchliffe {\it et al.}, 
  Phys.\ Rev.\ D {\bf 55}, 5520 (1997);
  B.~C.~Allanach, C.~G.~Lester, M.~A.~Parker and B.~R.~Webber,
  JHEP {\bf 0009}, 004 (2000)
  [arXiv:hep-ph/0007009].

\bibitem{Ellis:1982xd}
  J.~R.~Ellis and G.~G.~Ross,
  Phys.\ Lett.\ B {\bf 117}, 397 (1982);
  A.~Bartl, H.~Fraas and W.~Majerotto,
  Z.\ Phys.\ C {\bf 30}, 441 (1986);
  A.~Bartl, H.~Fraas and W.~Majerotto,
  Nucl.\ Phys.\ B {\bf 278}, 1 (1986).

\bibitem{Weiglein:2004hn}
  G.~Weiglein {\it et al.}  [LHC/LC Study Group],
  arXiv:hep-ph/0410364.


\bibitem{Fox:2002bu}
  P.~J.~Fox, A.~E.~Nelson and N.~Weiner,
  JHEP {\bf 0208}, 035 (2002)
  [arXiv:hep-ph/0206096].

\bibitem{Choi:2001ww}
  S.~Y.~Choi, J.~Kalinowski, G.~Moortgat-Pick and P.~M.~Zerwas,
  Eur.\ Phys.\ J.\ C {\bf 22}, 563 (2001).

\bibitem{CPT}
  S.~T.~Petcov,
  Phys.\ Lett.\ B {\bf 139}, 421 (1984);
  S.~M.~Bilenky, N.~P.~Nedelcheva and E.~K.~Khristova,
  Phys.\ Lett.\ B {\bf 161}, 397 (1985);
  G.~Moortgat-Pick and H.~Fraas,
  Eur.\ Phys.\ J.\ C {\bf 25}, 189 (2002);
  A.~Bartl {\it et al.}, 
  JHEP {\bf 0601}, 170 (2006)
  [arXiv:hep-ph/0510029].

\bibitem{Aguilar-Saavedra:2003hw}
  J.~A.~Aguilar-Saavedra and A.~M.~Teixeira,
  Nucl.\ Phys.\ B {\bf 675}, 70 (2003).

\bibitem{Choi:2003fs}
  S.~Y.~Choi and Y.~G.~Kim,
  Phys.\ Rev.\ D {\bf 69}, 015011 (2004);
  S.~Y.~Choi,
  Phys.\ Rev.\ D {\bf 69}, 096003 (2004);
  S.~Y.~Choi, B.~C.~Chung, J.~Kalinowski, Y.~G.~Kim and K.~Rolbiecki,
  arXiv:hep-ph/0504122.

\bibitem{Nojiri:1994it}
  M.~M.~Nojiri,
  Phys.\ Rev.\ D {\bf 51}, 6281 (1995).

\bibitem{Nojiri:1996fp}
  M.~M.~Nojiri, K.~Fujii and T.~Tsukamoto,
  Phys.\ Rev.\ D {\bf 54}, 6756 (1996).

\bibitem{Denner:1992vz}
  A.~Denner, H.~Eck, O.~Hahn and J.~Kublbeck,
  Nucl.\ Phys.\ B {\bf 387}, 467 (1992).

\bibitem{HELAS} 
  H. Murayama, I. Watanabe, and K. Hagiwara, 
  KEK Report 91-11 (1992).

\bibitem{Kawabata:1995th}
  S.~Kawabata,
  Comput.\ Phys.\ Commun.\  {\bf 88}, 309 (1995).

\end{thebibliography}
\end{document}